\documentclass[aoas,MSNbibl,nameyear,dvips]{arximspdf}
\usepackage{graphicx}
\doi{10.1214/11-AOAS518}
\volume{6}
\issue{2}
\pubyear{2012}
\firstpage{601}
\lastpage{624}

\makeatletter
\newcommand{\eqref}[1]{(\ref{#1})}
\newcommand{\cal}{\mathcal}
\makeatother

\begin{document}
\begin{frontmatter}

\title{Functional factor analysis for periodic remote sensing data\thanksref{TT1}}
\runtitle{Principal periodic factors}

\begin{aug}
\author[A]{\fnms{Chong} \snm{Liu}\ead[label=e1]{liuchong@math.bu.edu}},
\author[A]{\fnms{Surajit} \snm{Ray}\corref{}\ead[label=e2]{sray@bu.edu}},
\author[B]{\fnms{Giles} \snm{Hooker}\ead[label=e3]{gjh27@cornell.edu}}
\and
\author[C]{\fnms{Mark} \snm{Friedl}\ead[label=e4]{friedl@bu.edu}}
\runauthor{Liu, Ray, Hooker and Friedl}
\affiliation{Boston University, Boston University, Cornell University
and Boston~University}
\address[A]{C. Liu\\
S. Ray\\
Department of Mathematics and Statistics\\
Boston University\\
111 Cummington Street\\
Boston, Massachusetts 02215\\ USA\\
\printead{e1}\\
\hphantom{\textsc{E-mail: }}\printead*{e2}} %adresu isvedimo komanda
%gale!
\address[B]{G. Hooker\\
Department of Statistical Science and\\
\quad Department of Biological Statistics \\
\quad and Computational Biology\\
Cornell University\\
1186 Comstock Hall\\
Ithaca, New York 14853\\ USA\\
\printead{e3}}
\address[C]{M. Friedl\\
Department of Geography\\ \quad and Environment\\
Boston University\\
675 Commonwealth Ave\\
Boston, Massachusetts 02215\\ USA\\
\printead{e4}}
\end{aug}
\thankstext{TT1}{Supported by National Science Foundation Grants ATM-0934739, DEB-0813743 and NASA cooperative agreement NNX08AE61A.}

% HISTORY:
\received{\smonth{5} \syear{2011}}
\revised{\smonth{8} \syear{2011}}

% ABSTRACT
%
\begin{abstract}
We present a new approach to factor rotation for functional data. This is
achieved by rotating the functional principal components toward a predefined
space of periodic functions designed to decompose the total variation into
components that are nearly-periodic and nearly-aperiodic with a predefined
period. We show that the factor rotation can be obtained by calculation of
canonical correlations between appropriate spaces which make the methodology
computationally efficient. Moreover, we demonstrate that our proposed rotations
provide stable and interpretable results in the presence of highly complex
covariance. This work is motivated by the goal of finding interpretable sources
of variability in gridded time series of vegetation index measurements obtained
from remote sensing, and we demonstrate our methodology through an application
of factor rotation of this data.
\end{abstract}

% KEYWORDS
%
\begin{keyword}
\kwd{Factor rotation}
\kwd{variance decomposition}
\kwd{functional data analysis}
\kwd{covariance surface}
\kwd{remote sensing}
\kwd{principal periodic components}.
\end{keyword}

\end{frontmatter}

%s1 #&#
\section{Introduction}

The goal of factor rotation is to find interpretable directions
explaining the
covariance of the variables. In the case of classical multivariate data
interpretation of factors it is primarily carried out based on the
grouping of
factor loadings. However, these approaches are not always applicable to
collections of random functions. Instead, we propose an interpretable factor
rotation using a naturally predefined space of functions. The
motivating data
set for this paper consists of roughly weekly observations of vegetation
acquired from remote sensing at regular intervals for multiple years.
In this
case, the dominant seasonal cycle provides a natural choice for
dividing the
variation into nearly-periodic and nearly-aperiodic sources of
variation. More
generally, our approach facilitates understanding highly complex forms of
functional variation by dividing the total variation into two
orthogonal parts,
each of which may be explained by a smaller number of components with clear
interpretation. Besides achieving the desired interpretability, these
components are shown to be stable over the choice of the number of
factors and
can be obtained through computationally inexpensive steps.\looseness=-1

While a large amount of methodological development in functional data analysis
has been based on functional principal components analysis [\citet
{MSY07}] and
considerable theoretical attention devoted to its properties
[\citet{YMW05}, \citet{hall2006properties}, \citet
{li2010uniform}], little attention has been given to finding
rotations of the leading principal components to improve the interpretability
of variance components in fPCA. In this context, \citet{RamsaySilverman05}
propose a VARIMAX rotation, accomplished by evaluating derived principal
components on a fine grid; VARIMAX rotations yield components that have either
very high or very low values, effectively focusing variation on particular
regions of the functional domain. In many contexts, this can be useful---their
study of Canadian weather data neatly picks up the four seasons, for
example---but there is considerable further scope for alternative
notions of
interpretability to be developed. In particular, existing rotation
methods designed for multivariate data generally seek to emphasize particular
variables or observations, but do not attempt to account for the ordering
relations between variables that exist in functional data. We generally expect
the loading at one time point to be close to the loading at a nearby time
point. One way to achieve this is through smoothing penalties. Instead, we
define a rotation toward an interpretable reference subspace of functions.

\begin{figure}

\includegraphics{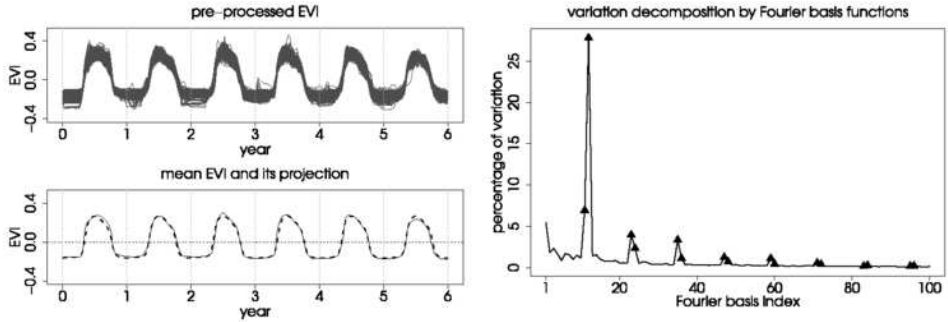}

\caption{Upper left: Preprocessed EVI data is obtained by first
smoothing raw
EVI observations with saturated Fourier basis expansion and the
penalty on
the second derivative and then the raw EVI fit is time-series
demeaned. Lower
left: The solid line is the mean of preprocessed EVI curves. The
dashed line
is the projection of the mean onto the subspace spanned by all Fourier basis
functions with annual period in the saturated basis system. Right: Percentage
of variation explained by Fourier basis functions. Preprocessed EVI curves
are projected onto each Fourier basis function. The variance of the
projection scores and its percentage of the total variance are
computed. The
Fourier basis index starts from $\sin(\omega t)$. The function $\sin
(K\omega t)$ has index
$2K - 1$ and $\cos(K\omega t)$ has index $2K$. The solid triangles highlight
the percentage score-variance associated with the annual Fourier basis which
correspond to index $11,12,23,24,35,36,\ldots,95,96$. The constant
basis is
not included in the calculation and index.}
\label{preproc}
\end{figure}

\begin{figure}

\includegraphics{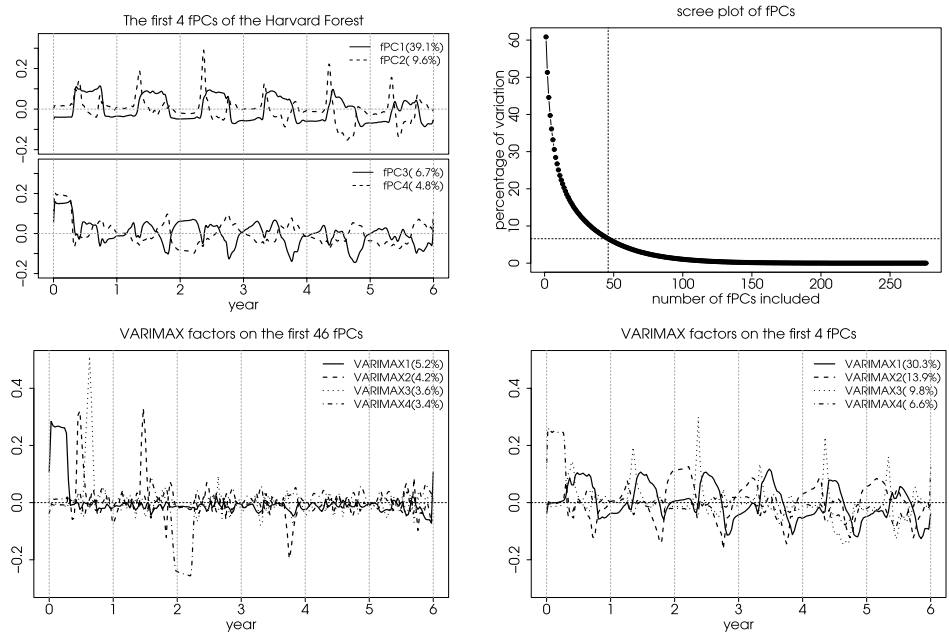}\vspace*{-3pt}

\caption{Upper left: the first 4 fPCs of the Harvard Forest data; Upper
right: the scree plot of the fPC. The vertical dashed line stands at
46 and
the horizontal dashed line shows the amount of variation not explained
by the
first 46 fPCs; Lower left: VARIMAX components derived by rotating the first
46 fPCs; Lower right: VARIMAX components derived by rotating the first 4
fPCs. Numbers in parentheses of the legend are percentage of variation
explained by each component.}
\label{pcandvarimax}\vspace*{-5pt}
\end{figure}

In the context of multi-year time series remote sensing data, the need for
methods to extract interpretable sources of variation is particularly acute.
The vegetation index considered in this paper consists of a 6-year time series
of remote sensing images acquired at 8-day intervals for a site in central
Massachusetts (see Section~\ref{subsecdata} and
Figure~\ref{preproc}).
These
data demonstrate a~highly complex functional covariance structure. To
illustrate,
in Figure~\ref{pcandvarimax} we present a scree plot of eigenvalues
for the
data set. This scree plot shows exponential decay in explained
variance, with
no evidence of the ``elbow'' that is frequently used to decide the
number of
eigenvalues to retain. Further, if we wished to explain 90\% of variation---a
frequently used criterion---over 30 components would need to be
retained, and
examination of the first few principal components suggests that interpretation
of these components is problematic (see Figure~\ref{pcandvarimax}),
consisting of both strong periodic structure as well as trends and isolated
features. Interpreting these sources of variation from this covariance
structure is challenging and common techniques such as VARIMAX
rotations (also
shown in Figure~\ref{pcandvarimax}) are clearly unhelpful in this
case. There
is, however, one clear and highly interpretable feature in the data: a strong
periodic signal. This is naturally expected due to the strong seasonal
forcing.\vadjust{\goodbreak}

%f1 #&#
%

%f2 #&#
%

Basing an interpretation around seasonality is both visually satisfying and
scientifically useful. Perhaps the most widely recognized feature of the
global climate (e.g., temperature, precipitation) and ecosystem (e.g.,
vegetation) data is seasonality [\citet{Hartmann1994}]. This can
be illustrated
by spectral decomposition of our data, shown in the right plot in Figure
\ref{preproc}, where annual variation dominates. Meanwhile, because climate
dynamics are produced by complex interactions among the Earth's oceans,
atmosphere, cryosphere, and land masses, the Earth's weather and
climate system,
and hence indicators of ecosystem, does not behave in a strictly periodic
fashion [\citet{holton1992introduction}]. Although sophisticated
models have been
developed for predicting climate-ecosystem dynamics, our understanding remains
incomplete.

\begin{figure}

\includegraphics{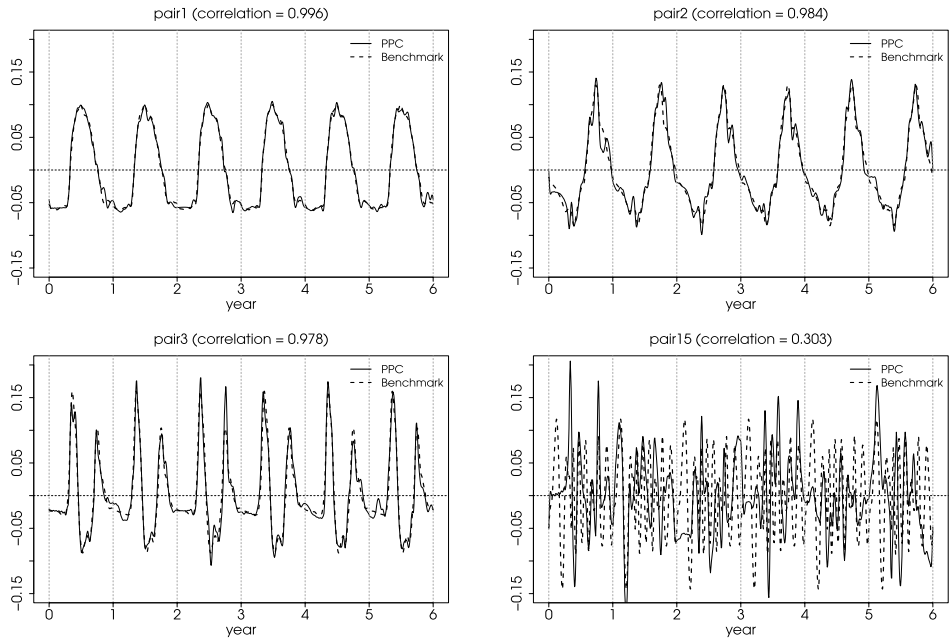}

\caption{PPC results on Harvard Forest data. PPCs are computed with 46 fPCs
of preprocessed EVI curves. The solid curves are PPCs $\xi_j$ and the dashed
curves are benchmarks~$\theta_j$ associated with~$\xi_j$. The
correlation is
computed as the standardized inner product between~$\theta_j$ and~$\xi_j$.
The pair index is ordered by the correlation.}
\label{ppc}
\end{figure}

The contribution of this paper is to provide a new factor rotation technique
that divides sources of variation into nearly-periodic and nearly-aperi\-odic
components. While strictly periodic components could be obtained
directly by
projecting onto a basis of periodic functions, the year-to-year
variation in
season timing requires us to retain somewhat more flexibility so as not to
overestimate the amount of nonseasonal variation. One approach to this would
be to undertake a registration procedure
[\citet{GerviniGasser04}, \citet{LiuMuller04}, \citet
{RamsaySilverman05}, \citet{KneipRamsay08}]. However,
the registration is ill-posed~and registration algorithms are computationally
expensive, particularly for large and complex data sets. Instead, we keep
within the framework of factor rotation and seek a rotation that
rotates the
largest sources of variation toward being periodic or a-periodic (see Figure
\ref{ppc}). This is accomplished via a canonical correlations transform
providing what we have labeled \textit{principal periodic components} (PPCs).

%f3 #&#
%

In comparing VARIMAX and PPC, we perform both rotations on a sequence
number of
fPCs and compute the change in $L^2$ sense between the first rotated components
derived from two consecutive numbers of fPCs. The $L^2$ change of PPC rotation
is much smaller and more stable compared to VARIMAX rotation,
suggesting PPC's
robustness with respect to the number of fPCs used in rotation.

Simulation studies also show that PPCs perform very well in detecting periodic
variation in the following two cases: (i) amount of periodic variation
increases from 0 to a level only comparable to other source of
variation where
fPCs react slowly to the increasing periodic variation; (ii) total
variation is
dominated by increasing amount of high frequency disturbances where
fPCs are
quickly contaminated by disturbances and PPCs still capture the
periodic source
of variation.\vadjust{\goodbreak}

To better understand the rotation and the relation between PPCs and the
space of functions with strict annual cycle, we develop a heuristic
test of whether the first PPC lies in that space. In the test, we
create a set of curves under the null hypothesis as close to the
original data as possible by either replacing the first PPC by its
associated benchmark, or inflating the nearly-periodic component while
controlling for Kullback--Leibler divergence of the sample functional
covariance to the null functional covariance. The test on our
motivating data set rejects the null hypothesis, suggesting that no
strict annual variation is presented in the space spanned by PPCs.

A further aspect of the data is that it is gridded in a regular spatial
distribution. This induces both spatial correlation as well as effects
due to (unobserved) geographic and environmental factors. Our use of
rotations will allow the effect of these structures to be empirically
investigated in terms of both variation in cyclic ecological factors
and in longer-term trends. Our functional data analysis approach
differs from techniques using empirical orthogonal functions (EOFs) in
spatio-temporal analysis [e.g., in \citet{eof}] in considering
observations as functions of time rather than of space. We believe that
this approach is appropriate to the task of separating cyclical from
other trends. We note that a similar rotation of EOFs toward a subspace
of functions describing landscape features or other geographic
gradients could be developed along similar lines to PPCs, but this is
beyond the scope of the current paper.

The remainder of this paper is arranged as follows. In Section
\ref{secmotivating} we will show a motivating example in which we
carry out
smoothing and functional principal component analysis and demonstrate the
motivation for PPCs. In Section~\ref{secppc} we introduce the
framework of
PPC and its results on our remote sensing data. Results of a simulation
study are
presented in Section~\ref{secsim} that illustrate the sensitivity and
robustness of PPC in identifying periodic variation. Details of further
simulation experiments concerning the power and size of the proposed
tests are given in
the supplemental article [\citet{ppcsup}]. We end with some
concluding remarks and discussion of future research.

%s2 #&#
\section{A motivating example}\label{secmotivating}

%s2.1 #&#
\subsection{The data set}\label{subsecdata}

The data set used for this work consists of time series of remotely sensed
images acquired over a site in central Massachusetts. Specifically, we used
surface spectral reflectance measurements from the Moderate Resolution Imaging
Spectroradiometer (MODIS) onboard NASA's Terra and Aqua satellites.
Data from
MODIS were extracted for a 25 by 25 pixel window (covering an area of
$\approx$
134 km$^2$) centered over the Harvard Forest Long Term Experimental Research
site in Petersham, MA. This site is characteristic of mid-latitude temperate
forests and is dominated by deciduous tree and understory species that exhibit
strong seasonal variation in phenology.
Data are provided at 8-day intervals (46 data
points per year) for the period from January 1, 2001 to December 31,
2006. The
spatial resolution of the data is 500-m on the ground.

Using MODIS surface spectral reflectances in the blue, red and near-infrared
(NIR) wavelengths, we computed a quantity known as the ``enhanced vegetation
index'' [EVI; \citet{huete2002overview}]:
\[
2.5\times\frac{\mathit{NIR}-\mathit{RED}}{\mathit{NIR} + C_1\times \mathit{Red}-C_2\times \mathit{Blue} + L},
\]
where NIR, Red and Blue are reflectances of the corresponding bands
re\-corded by
MODIS and $C_1$, $C_2$ and $L$ are constant coefficients. The EVI exploits
spectral reflectance properties of live vegetation, yielding an index that
scales from $-$1 to 1 that is widely used for monitoring seasonal
dynamics in
vegetation. Because EVI data are sensitive to the presence of snow and include
noise and missing values caused by clouds, the data were preprocessed prior
to analysis to remove noise and fill gaps following the procedure
described by
\citet{zhang2006global}.
In the supplemental article [\citet{ppcsup}],
we provide a detailed account of this preprocessing and criteria for
excluding pixels with large
blocks of missing observations.\vadjust{\goodbreak}

The final data set consisted 276 EVI
time series values for each of 423 pixels (excluding pixels with problematic
observations), that is, 423 replicated curves, with each replication
corresponding
to a pixel in the area of interest. The regular spatial and temporal sampling
of EVI data makes functional data analysis a useful framework for exploring
variation among curves and facilitating the study of change in variation.

%s2.2 #&#
\subsection{Smoothing of EVI}
Denote the discrete observation at pixel $i$ and time $t_{ij}$ by
$Y_{ij}$. We
consider the following additive error model:
\begin{eqnarray*}
Y_{ij} = x_i(t_{ij}) + e_{ij}, \qquad1\leq i\leq N \mbox{ and }
1\leq j\leq n_i,
\end{eqnarray*}
where $x_i(t)$'s are the true realizations of the underlying random
growing process $X(t)$ and $e_{ij}$'s are errors. To estimate $x_i(t)$, we
choose a regularization approach based on basis expansion.
Specifically, we fit
our data with the saturated Fourier basis and explicitly penalize the total
curvature. The Fourier basis is numerically convenient for our
purposes; experiments with
alternative B-spline bases indicated that our results are insensitive
to this choice.
The smoothing parameter is chosen to minimize the sum of generalized
cross validation scores over all curves. This can be implemented in R
[\citet{R}]
using the FDA package [\citet{fdaR}]. Let ${\hat x}_i(t)$ denote
the fitted curve. Then,
we further process this raw fitting by removing from each ${\hat
x}_i(t)$ its
time-series average. Then, we obtain the demeaned curve $z_i(t)$ as
\[
z_i(t) = {\hat x}_i(t) - \frac{1}{T}\int^T_0 {\hat x}_i(t)\,dt, \qquad
1\leq i\leq N.
\]

This demeaning process removes vertical variation and avoids defining
it as either annual or nonannual. The centering removes heterogeneity
in the overall growing level and allows us to focus on nonconstant
modes of
variation; see further discussion in Section~\ref{secppc}.
The
pre-smoothed and demeaned EVI curves of the Harvard Forest data are
shown in the upper
left plot in Figure~\ref{preproc}. While many methods have been
developed to analyze the
features and structure of the mean shape, in this paper we are
interested in
changes in vegetation dynamics manifested in terms of variance.

We decompose the total variation among the EVI curves by projecting EVI curves
to the saturated Fourier basis system. This decomposition shows that variation
explained by the annual Fourier basis function is a dominating source
of variation
in our example, as shown in the right plot in Figure~\ref{preproc}.
The data here are defined on a grid of observations
taken every 8 days and thus could be considered a very high-dimensional
multivariate data set. We have chosen to view these data as functional
due to
the underlying smooth greening process that they record, and because it
facilitates the definition of periodicity which we employ to define a factor
rotation below.

%%%%%%%%%%%%%%%%%%%%%%%%%%%%%%%%%%%%%%%%%%%%%%%%%%%%%%%%%%%%%%%%%%%%%%%%%%%%%%%
%%%%%%% Principal Component Analysis of functional objects
%%%%%%%%%%%%%%%%%%%%%%%%%%%%%%%%%%%%%%%%%%%%%%%%%%%%%%%%%%%%%%%%%%%%%%%%%%%%%%%
%s2.3 #&#
\subsection{Functional principal component analysis}

Functional principal component analysis (fPCA) is a well studied
research area. It provides a way to extract the major mode of variation
among curves and our proposed PPC is based on and motivated by fPCA. To
introduce and fix notation
for description of PPC in later sections, we give a brief review on
fPCA. More references on fPCA can be found in
\citet{RamsaySilverman02}, \citet{YMW05}
and \citet{MSY07}. In particular, we look for a set of normalized
and orthogonal functions $\gamma_j(t)$ such that the projection of all
EVI curves onto each specific $\gamma_j(t)$ has the largest
variability. These $\gamma_j(t)$'s are called the functional principal
components (fPCs).
Formally, suppose we have $N$ smoothed and time-series demeaned EVI curves
$z_i(t)$, $1\leq i\leq N$. The sample cross-section mean process is
${\hat\mu}(t) = N^{-1}\sum_i z_i(t)$. Then the cross-section demeaned
curve is obtained as $\tilde{z}_i(t) = z_i(t) - {\hat\mu}(t)$.
$\gamma
_j(t)$ is chosen to maximize $N^{-1}\sum_i(\int\gamma_j(t)\tilde
{z}_i(t)\,dt)^2$ subject to the constraints that $\int{\gamma
_j(t)\gamma
_k(t)\,dt} = \delta_{jk}$ where $\delta_{jk}$ is the Kronecker delta.
Given the estimated covariance kernel $\Omega(s,t) = N^{-1}\sum
_{i=1}^N\tilde{z}_i(s)\tilde{z}_i(t)$, each fPC, $\gamma_j(t)$,
satisfies the eigen-equation $\int\Omega(s,t)\gamma_j(t)\,dt = \lambda
_j\gamma_j(s)$, where~$\lambda_j$ is the associated eigenvalue. By
writing $\gamma_j(t)$ in expansion of basis functions, this problem can
be reduced to the computation of matrix eigenvalues and eigenvectors.
Here we have pre-smoothed the data and applied a~principle-components
decomposition without additional penalty. fPCA can also be employed
along with smoothing methods [\citet{Silverman96}] or by directly
smoothing the covariance surface [\citet{YMW05}]. The methods
developed below are
applicable for an fPCA decomposition, irrespective of the method
employed to derive it.

In order to explore the variation in EVI curves, we apply the standard fPCA
techniques on Harvard Forest data. The first 4 fPCs of
Harvard Forest are plotted in Figure~\ref{pcandvarimax} where each of
the four fPCs contains some level of annual
periodicity and pick up features of EVI variation at different times of year.
For example, the first PC shows that the contrast of EVI between summer and
winter is the most distinct feature that characterizes the vegetation growing
in this area, however, with a decreasing trend suggesting the contrast between
summer and winter has changed over the 6 years. The second fPC has a
sharp peak
roughly at the start of each growing season combined with noticeable dips
during years 5 and~6. Due to the existence of the two negative bumps,
it is hard
to interpret the second fPC as the effect of growing season onset.
A third fPC emphasizes the ending of growing
season, characterizing variation in the timing of vegetation browning.
However, as these fPCs are combined with nonannual signal, they are not
designed to distinguish between annual and nonannual sources of variation.
In Section
\ref{secppc} we discuss the appropriate rotation of fPCs to aid
interpretation by separating annual and nonannual sources of variation.
But first we
discuss one widely used technique of rotation for functional data---the
VARIMAX rotation.

%s2.4 #&#
\subsection{VARIMAX rotation}

VARIMAX is a widely used orthonormal
transformation in multivariate analysis which can make multivariate principal
components more interpretable. The functional VARIMAX rotation borrows readily
the concept of multivariate VARIMAX rotation. Suppose we retain the
first $M$
fPCs and the subspace spanned by these $M$ fPCs is denoted by
${\boldsymbol\Gamma}_M$.
We use ${\boldsymbol\gamma}$ to refer to the vector valued function
$(\gamma_1,\ldots,\gamma_M)'$. Let $\mathbf B$ be a $M\times n$
evaluation matrix
of $\boldsymbol\gamma$ where $\mathbf{B}_{ij} = \gamma_i(t_j)$,
$1\leq
j\leq n$. Given
an orthonormal matrix $\mathbf T$, ${\boldsymbol\nu} = \mathbf
{T}{\boldsymbol\gamma}$ gives us a new
set of orthonormal functions. The evaluation matrix at the same $t_j$'s
of the
rotated functions~$\boldsymbol\nu$ is given by $\mathbf A = \mathbf{TB}$. Denote
the $ij$th entry of
matrix A by $a_{ij}$. Then the VARIMAX strategy for choosing the orthonormal
rotation matrix $\mathbf T$ is to maximize the variation of $a_{ij}^2$
over all
values of $i$ and $j$.

The solution to the above maximization problem will encourage values $a_{ij}$
to be either strongly positive, near zero, or strongly negative. This rotation
tends to cluster information and make the components of variation
easier to
interpret. \mbox{VARIMAX} rotation on the first 46 fPCs
and on the first 4 fPCs are shown in the two lower plots in Figure~\ref{pcandvarimax}. If using only 4
fPCs, we do not have sufficient flexibility to provide improved interpretation.
By contrast, using 46 fPCs provides so much concentration on individual
time points
that any natural interpretation is lost.

The rotation described here can be generalized to describe a rotation
of principal components to find directions that lie close to an
interpretable reference subspace. In a multivariate context, this
amounts to finding a~rotation of factors $\Gamma$ so that the leading
components lie close to a subspace spanned by the columns of a matrix
$F_P$, assumed to have interpretable relevance for the application at
hand, and the mathematical development below can be read in an entirely
multivariate context. It more generally applies to observations taking
values on any Hilbert space. While the space of periodic functions is
clearly relevant for our application, the choice of subspace is
context-specific.

%%%%%%%%%%%%%%%%%%%%%%%%%%%%%%%%%%%%%%%%%%%%%%%%%%%%%%%%%%%%%%%%%%%%%%%%%%%%%%%
%%%%%%% Factor Rotation using functional canonical correlation
%%%%%%%%%%%%%%%%%%%%%%%%%%%%%%%%%%%%%%%%%%%%%%%%%%%%%%%%%%%%%%%%%%%%%%%%%%%%%%%
%s3 #&#
\section{Principal periodic component (PPC)} \label{secppc}
The VARIMAX rotation does not achieve our goal of separating annual and
nonannual variation since its objective function is not designed to do
so. We
need to explicitly define an objective function which can extract annual
variation. One natural way to do this is to order the rotated fPCs by their
levels of annual periodicity. To measure annual periodicity, we will first
define benchmarks which have strict annual periodicity. Then we compute the
closeness between rotated fPCs and corresponding benchmarks and this computed
closeness serves as the measure of annual periodicity of the rotated
fPCs. Refer
to Fourier basis functions with annual period as $f_k$, $1\leq k\leq
P$, the
vector of them as $\mathbf f$ and the space spanned by them as $\mathbf
{F}_P$. Hence,
$\mathbf{F}_P$ is a space of functions with annual periodicity up to a certain
frequency determined by $P$.
$P$ is limited to
the set of periodic Fourier coefficients used to smooth the data. More
generally, $P$ can be set to $N$---allowing the interpolation of any $N$ points
that lie in a strictly periodic subspace.
We construct
our benchmarks as the linear combination of $f_k$'s. Then
benchmarks are in $\mathbf{F}_P$ and thus have exactly annual periodicity.
Intuitively, we can consider $\boldsymbol\gamma$ and $\mathbf f$ as two
frames of their own
spaces ${\boldsymbol\Gamma}_M$ and $\mathbf{F}_P$. We can rotate the
two frames and align
them in the same direction as much as possible. If ${\boldsymbol\Gamma
}_M$ contains
direction which is exactly annual, then we will align the two spaces at least
in one direction. The closeness between the rotated fPC and associated
benchmark is computed as their standardized inner-product.

%s3.1 #&#
\subsection{Principal periodic component framework}

In this section we give a~mathematical description of the PPC methodology.
Recall that ${\boldsymbol\gamma}$ is a~$M$ dimensional vector of fPCs
obtained from
time-series demeaned curves and~$\mathbf f$ is a $P$ dimensional vector
of Fourier
basis functions with annual period. Define $\boldsymbol\Sigma_{\gamma
f} =
\langle {\boldsymbol\gamma}, \mathbf{f}\rangle $, where $\langle \cdot,\cdot\rangle $ is the
inner-product in $L^2$
space and the $ik$th entry of $\boldsymbol\Sigma_{\gamma f}$ is given
by $\langle \gamma_i,
f_k\rangle $. We compute the singular value decomposition $\boldsymbol\Sigma
_{\gamma f} =
{\hat{\mathbf U}'}\mathbf{W}{\hat{\mathbf V}}$ and denote the $j$th row
of ${\hat{\mathbf U}}$ by
${\hat{\mathbf u}_j}'$ and the $j$th row of~${\hat{\mathbf V}}$ by~${\hat{\mathbf v}_j}'$. The
PPCs and associated benchmarks are then defined as follows,
%e1 #&#
%
\begin{eqnarray}
\label{ppcbmkdef}
\xi_j = {\hat{\mathbf u}_j'}{\boldsymbol\gamma}  \quad \mbox{and}
\quad
\theta_j
= {\hat{\mathbf v}_j'}\mathbf{f}, \qquad j = 1,2,\ldots,\min(M,P).
\end{eqnarray}
In the above definition, we call $\xi_j$ the $j$th PPC and $\theta_j$ the
associated benchmark of~$\xi_j$.

In order to derive these estimates, denote any rotation on $\boldsymbol
\gamma$ by $\mathbf
U$ with~$\mathbf u'_j$ being the $j$th row of $\mathbf U$, and any
rotation on $\mathbf f$
by $\mathbf V$ with $\mathbf v_j'$ being the $j$th row of $\mathbf V$.
Let $\xi_j^0 = \mathbf
u'_j\boldsymbol\gamma$ and $\theta_j^0 = \mathbf v'_j\mathbf f$. Then
$\xi_j^0$ is the $j$th
rotated fPC and~$\theta_j^0$ is a function with annual cycle. We
define the
closeness measure of the pair~$\xi_j^0$ and~$\theta_j^0$ as the angle between
them,
%e2 #&#
%
\begin{eqnarray}
\label{rho}
\rho_j = \rho(\xi_j^0, \theta_j^0) = \frac{\langle {\xi}_j^0,
{\theta}_j^0\rangle }{\|{\xi}_j^0\|\|{\theta}_j^0\|} =
\frac{\langle {\mathbf u_j'}{{\boldsymbol\gamma}}, {\mathbf v_j'}\mathbf{f}\rangle }
{\|{\mathbf u_j'}{{\boldsymbol\gamma}}\|\|{\mathbf v_j'}\mathbf{f}\|}.
\end{eqnarray}
Given this closeness measure, we solve the following optimization
problem for $j = 1,2,\ldots,\min(M,P)$,
%e3 #&#
%
\begin{eqnarray}\label{ppcformulation}
({\hat{\mathbf u}_j}, {\hat{\mathbf v}_j}) = \operatorname{\arg\max
}\limits_{{\mathbf u_j, \mathbf v_j}} \rho(\xi_j^0, \theta_j^0)
= \operatorname{\arg\max}\limits_{{\mathbf u_j,\mathbf v_j}} \frac
{{\mathbf
u_j'}{{\boldsymbol\Sigma}_{\gamma{\it f}}}{\mathbf v_j}}
{{\mathbf u_j'}{\boldsymbol\Sigma}_{\gamma\gamma}{\mathbf u_j}\cdot
{\mathbf v_j'}
{\boldsymbol\Sigma}_{{\it f}{\it f}}{\mathbf v_j}},
\end{eqnarray}
subject to
$ \langle {\xi}_j^0,{\xi}_k^0\rangle  = \delta_{jk}$, $\langle {\theta}_j^0,{\theta
}_k^0\rangle  =
\delta_{jk}$, $\langle {\xi}_j^0,{\theta}_k^0\rangle  = 0,$
where the $ik$th entry of~${\boldsymbol\Sigma}_{\gamma f}$,
${\boldsymbol\Sigma}_{\gamma\gamma}$ and ${\boldsymbol\Sigma
}_{ff}$ are
given by
$\langle {\gamma}_i, {f}_k\rangle $, $\langle {\gamma}_i, {\gamma}_k\rangle $ and $\langle {f}_i,
{f}_k\rangle $,
respectively.

We observe that the objective in $\eqref{ppcformulation}$ has the same
form as
multivariate canonical correlation analysis (CCA) where random
variables are
replaced by functions. However, the sampling properties of the PPC
rotation differ from CCA in
that the frame of our reference subspace, $\mathbf f$, is deterministic
while fPCs
$\boldsymbol\gamma$ is random where randomness comes from sampling variation.
See \citet{mardia1980multivariate} for an overview of CCA; in the
functional analysis context see \citet{leurgans1993canonical} and
\citet{he2003functional}. According to the CCA results, we have
the following
solution:
%e4 #&#
%e5 #&#
%
\begin{eqnarray}
\label{ppcraw}
\begin{tabular}{p{300pt}}
 \mbox{${\hat{\mathbf u}_j} \mbox{ is proportional to the } j\mbox{th
eigenvector of }
\boldsymbol\Sigma_{\gamma\gamma}^{-1}
\boldsymbol\Sigma_{\gamma f}\boldsymbol\Sigma_{ff}^{-1}{\boldsymbol
\Sigma'_{\gamma f}}
\mbox{ and}$}   \mbox{${\hat{\mathbf u}_j'}{\boldsymbol\Sigma_{\gamma\gamma
}}{\hat{\mathbf u}_j} = 1$,}
\end{tabular}\\
\label{bmkraw}
\begin{tabular}{p{300pt}}
 \mbox{$
 {\hat{\mathbf v}_j} \mbox{ is proportional to the } j\mbox{th
eigenvector of }
\boldsymbol\Sigma_{ff}^{-1}{\boldsymbol\Sigma'_{\gamma f}}
\boldsymbol\Sigma_{\gamma\gamma}^{-1}\boldsymbol\Sigma_{\gamma f}
\mbox{ and}$}   \mbox{${\hat{\mathbf v}_j'}{\boldsymbol\Sigma_{ff}}{\hat
{\mathbf
v}_j} = 1$.}
\end{tabular}
\end{eqnarray}
Due to the orthogonality of fPCs and the Fourier basis system, we have
${\boldsymbol\Sigma}_{\gamma\gamma} = \mathbf{I}$ and
${\boldsymbol
\Sigma}_{ff} = \mathbf{I}$. These
two identities reduce \eqref{ppcraw} and \eqref{bmkraw} to the eigenanalysis
of ${\boldsymbol\Sigma}_{\gamma f}$ and the results in \eqref
{ppcbmkdef} follows. In
\eqref{ppcbmkdef}, $\hat{\mathbf U}$ and $\hat{\mathbf V}$ are two
orthogonal rotation
matrices on $\boldsymbol\gamma$ and $\mathbf f$, respectively. In a
more general context,
(\ref{ppcraw}) and (\ref{bmkraw}) can be employed if, for example, the space
$\mathbf{F}_P$ is not parameterized by an orthogonal basis.

In this context, the motivation for removing the time series mean of the
observations as described in Section~\ref{secmotivating} becomes
apparent. We
have not defined variation in terms of a constant vertical shift as being
either periodic or aperiodic in nature. Demeaning the observations ensures
that there is zero variation in this direction and, hence, all the
computed fPCs
will also integrate to zero. Had this step not been carried out, the constant
shift would have been conflated with both periodic and aperiodic
sources. If
this constant source of variation were defined as periodic, a constant function
could be added to the space $\mathbf{F}_P$.

%s3.2 #&#
\subsection{PPC results on Harvard forest data} \label{secppcHF}

We now apply our PPC methodology on the Harvard Forest data. In the Harvard
Forest data, periodicity is set to be annual and thus we have 46
Fourier basis
functions with annual period. Thus, $P = 46$ and the space spanned by these
functions is $\mathbf{F}_{46}$. We set $M = P = 46$ in our calculation
in order to
get pairwise match between the PPCs and benchmarks.
${\boldsymbol\Gamma}_{46}$ accounts for 93.4\% of total variation. The
robustness of PPC
computation with respect to the choice of $M$ is further discussed in Section
\ref{stability}. A selection of four pairs of PPCs and associated benchmarks
with decreasing correlations are shown in Figure~\ref{ppc}.

The first PPC suggests the most important annual variation is the contrast
between summer and winter. The second PPC has the effect of shifting summer
forward or backward in time, while the third PPC corresponds to combined
effect of growing season length and summer maximum EVI. These leading PPCs
demonstrate modes of variation which are most likely to repeat every
year. From
an ecological perspective, these sources of variance are of critical importance
because they reflect signatures of climate variability in ecosystem processes.
Thus, PPCs provide a tool for characterizing and understanding how subtle
changes in climate, such as shifts in the timing of seasons, are affecting
ecosystems [\citet{Parmesan}, \citet{Piao}].

Note that benchmarks are always exactly annual and the correlation
between PPCs
and their benchmarks decreases as we extract more PPCs. We thus
construct a set of orthonormal functions which are ordered by their
level of
annual periodicity. This shows that the amount of annual variation
contained in
PPCs decreases as the index increases. A trade-off in defining which
components should be
denoted ``periodic'' is detailed in Section~\ref{VarDecompsec}

%s3.3 #&#
\subsection{Stability of PPC directions}\label{stability}

Choosing the number of fPC components, $M$ is a statistically challenging
task. This number depends on several factors, including strength of
signals, the
choice of smoothing parameter, and sampling error as well as the
choice of fPCA methodology.
An ideal factor rotation should be insensitive to the number of factors
chosen. This is particularly important when there are many small
components of
variation and the number of components selected can be unstable. In the VARIMAX
rotation the interpretation of rotated components is very sensitive to
$M$. On
the other hand, the PPCs provide a natural framework to achieve this
goal when
the principal sources of variation are periodic in nature. This is
achieved due to the use of a well-defined reference subspace, thereby
stabilizing the choice of ``interesting'' directions.

\begin{figure}

\includegraphics{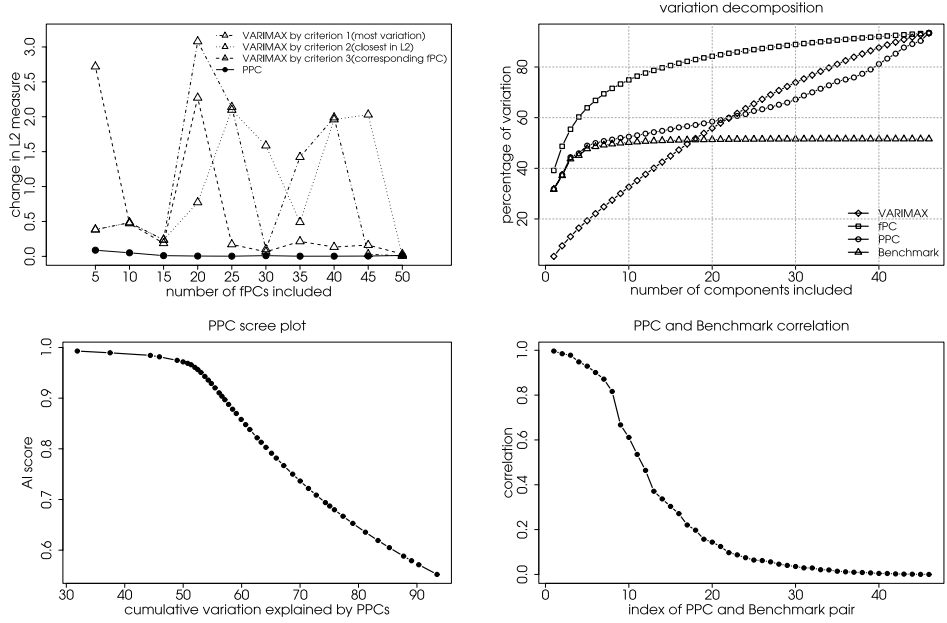}\vspace*{-5pt}

\caption{Upper left: $L^2$ difference between rotated components
derived on consecutive values of $M$. The
horizontal axis represents the value of $M$, the number of fPCs used
in rotations. For a given $M$, the
corresponding value on the vertical axis measures the $L^2$ difference
between components obtained by
rotating $M$ and $M - 5$ fPCs. Three different dashed lines correspond
to three definitions of the first
VARIMAX component. The solid line corresponds to the PPC. Upper right:
Percentage of variation explained.
Diamonds are the cumulative variation explained by VARIMAX components.
Squares are the cumulative variation
explained by fPCs. Circles are the cumulative variation explained by
PPCs. Triangles are the cumulative variation
explained by benchmarks. Lower left: PPC scree plot, computed as the
amount of cumulative variation
explained by benchmarks as a proportion of cumulative variation
explained by PPCs; Lower right:
Correlation between PPCs and benchmarks.}
\label{scree}\vspace*{-4pt}
\end{figure}

We explored the stability of the leading rotated component for a range of
choices for $M$---the number of fPCs we rotate---from~5 to 50 in increments
of~5. In these data, the first VARIMAX component was highly unstable,
while the
first PPC remained stable and retained most of its interpretation for
the whole
range of $M$ (see Figure~\ref{scree}). Here we define the first important
VARIMAX component in any of three ways: (i) the component which
accounts for
the most variation, (ii) the component of $M$ fPCs that is closest to the
first VARIMAX direction derived with $M - 5$ fPCs in the $L^2$ sense,
and (iii) the
component closest to the first fPC. To summarize the stability of these
rotations, we explored the $L^2$ difference between components rotated
using~$M$
and $M + 5$ fPCs under each of the three VARIMAX definitions above and using
the first PPC component. The $L^2$ differences on PPC rotation are highly
stable, whereas the measure for all of the VARIMAX rotations shows
large change
in both directions.

%f4 #&#
%

%s3.4 #&#
\subsection{Variation decomposition} \label{VarDecompsec}

We demonstrate the variation decomposition using two sets of rotations, one
being the standard VARIMAX rotation and the other being the PPC rotation
described in this paper. For comparing the two techniques we define component
scores as EVI curves projected on the set of orthogonal functions in\vadjust{\goodbreak}
which we
are interested. Denote the VARIMAX components based on 46 fPCs by $\nu
_j$. Then
we have
\begin{eqnarray*}
s^{\gamma}_{ij} &=& \int_T\tilde{z}_i(t)\gamma_j(t)\,dt   \quad
\mbox{and}\quad
\lambda^{\gamma}_j = \frac{1}{N-1}\sum_{i = 1}^N(s^{\gamma
}_{ij})^2,\\[-3pt]
s^{\xi}_{ij} &=& \int_T\tilde{z}_i(t)\xi_j(t)\,dt \quad\mbox{and}\quad
\lambda^{\xi}_j = \frac{1}{N-1}\sum_{i = 1}^N(s^{\xi}_{ij})^2,\\[-3pt]
s^{\theta}_{ij} &=& \int_T\tilde{z}_i(t)\theta_j(t)\,dt \quad\mbox{and}\quad
\lambda^{\theta}_j = \frac{1}{N-1}\sum_{i = 1}^N(s^{\theta
}_{ij})^2,\\[-3pt]
s^{\nu}_{ij} &=& \int_T\tilde{z}_i(t)\nu_j(t)\,dt \quad\mbox{and}\quad
\lambda^{\nu}_j = \frac{1}{N-1}\sum_{i = 1}^N(s^{\nu}_{ij})^2.\vadjust{\goodbreak}
\end{eqnarray*}
The cumulative sum of $\lambda^{\gamma}_j, \lambda^{\xi}_j,
\lambda^{\theta}_j$
and $\lambda^{\nu}_j$ are plotted in the upper right plot in Figure
\ref{scree}. The VARIMAX decomposition tends to produce equal decomposition,
indicated by the low curvature of its cumulative sum. The fPCs
decompose the
total variation by their decreasing abilities to explain variation, producing
the concave feature seen in its cumulative sum. Variation explained by the
benchmarks goes flat, suggesting the annual variation represented by benchmarks
with low correlation tends to be orthogonal to ${\boldsymbol\Gamma
}_{46}$. The
increasing gap from the left to the right between PPC decomposition and
benchmark decomposition reflects the decreasing ability to line up the rotated
frames of ${\boldsymbol\Gamma}_{46}$ and $\mathbf{F}_{46}$.

%s3.5 #&#
\subsection{Nearly-annual and nonannual trade-off} \label{sectradeoff}

In this subsection we develop an ad hoc methodology of choosing PPCs as
nearly-annual, in order to separate annual variation from nonannual variation.
Since the level of annual periodicity decreases, it suffices to find a cut-off
position and include all PPCs before the cutoff as nearly-annual and
all PPCs
after the cutoff as nonannual. To this end, we measure the cumulative amount
of variation explained by benchmarks as a proportion of cumulative variation
explained by PPCs and call it annual information (AI). Specifically, we define
\[
AI_j = \frac{\sum_{k=1}^j\lambda^\theta_k}{\sum_{k=1}^j\lambda
^\xi_k}.
\]
AI scores show an elbow around 8 PPCs (see the lower left plot in Figure~\ref{scree}). This elbow suggests a possible position to cutoff. This position
is further supported by the plot of correlation between PPCs and benchmarks
where a sudden drop is observed around 8 PPCs.
In the supplemental article [\citet{ppcsup}] we detail a
simulation study investigating the efficacy of AI as a visual diagnostic
where we demonstrate that the appropriate number of PPC's is selected
with high probability.

%s3.6 #&#
\subsection{Application of PPC}

PPCs are modes of variation which are ordered by their level of annual
periodicity. Since PPCs are generated by orthogonally rotating the
fPCs, PPCs
form another empirical orthogonal basis which can be used to decompose EVI
curves. Moreover, if we project EVI curves onto PPCs and fPCs, the
approximation by PPCs is as good as the approximation by fPCs. However,
we can
further decompose EVI curves into nearly-annual and nonannual components.
Suppose $P > M$ and thus we have $M$ PPCs. If we have $K$ fPCs in
total, then
we have the following decomposition:
%e6 #&#
%
\begin{eqnarray}\label{decompose}
z_i(t) = {\hat\mu}(t) + \sum_{j=1}^Js^{\xi}_{ij}\xi_j(t) + \sum
_{j=J+1}^{M}s^{\xi}_{ij}\xi_j(t) +
\sum_{j=M+1}^{K}s^{\gamma}_{ij}\gamma_j(t), \nonumber
\\[-8pt]
\\[-8pt]
\eqntext{1\leq i\leq N.}
\end{eqnarray}

\begin{figure}

\includegraphics{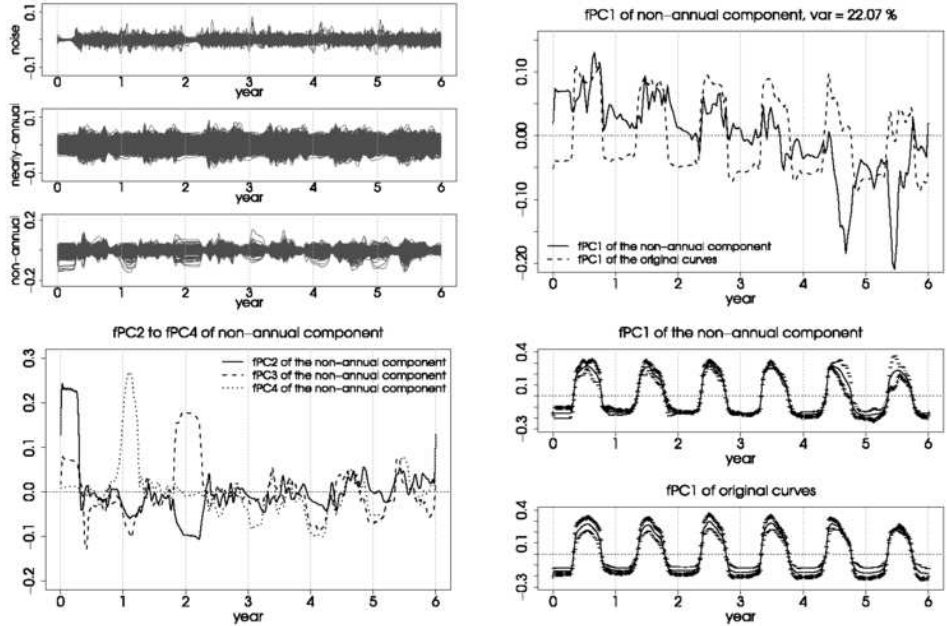}

\caption{Upper left: Decomposition of signals. The top panel is
noise which is removed when we retain the first 46 fPCs. The middle
panel is
the nearly-annual component and the bottom panel is the nonannual component.
Cutoff between nearly-annual and nonannual is chosen at 8 first PPCs.
Upper right:
The first fPCs. The dashed curve is the first fPC of the
original data. The solid curve is the first fPC of the nonannual
component. Lower
left: The second, the third and the fourth fPCs of the nonannual
component. Lower right:
Interpretation of the first fPCs of original data and nonannual
component. Solid curves are the
mean. In the upper panel, plus signs are mean curve plus multiple of the
first nonannual fPC and minus signs are mean curve minus multiple of
the first
nonannual fPC. In the lower panel, plus and minus signs are the
multiple of the first fPC of original data away from the
mean curve.}\label{non-annual}
\end{figure}

The first term on the right-hand side of \eqref{decompose} is the
sample mean
function. The second and the third terms are the nearly-annual
component and
nonannual component we determined in the last subsection. Note $J$ in
\eqref{decompose} is the number of PPCs we determined as nearly-annual.
For the
Harvard Forest data, $J$ is taken as 8 based on the AI elbow and the correlation
criterion described in Section~\ref{sectradeoff}. The last term in
\eqref{decompose} is the contribution of fPCs associated with very small
eigenvalues, which are removed when we truncate to a certain percentage of
variation. These are retained in conducting the simulation studies
below. The
decomposition result is shown in Figure~\ref{non-annual}. This decomposition
helps us reconstruct original EVI curves with restoring annual
information as
our priority.

%f5 #&#
%

%f6 #&#
%
\begin{figure}

\includegraphics{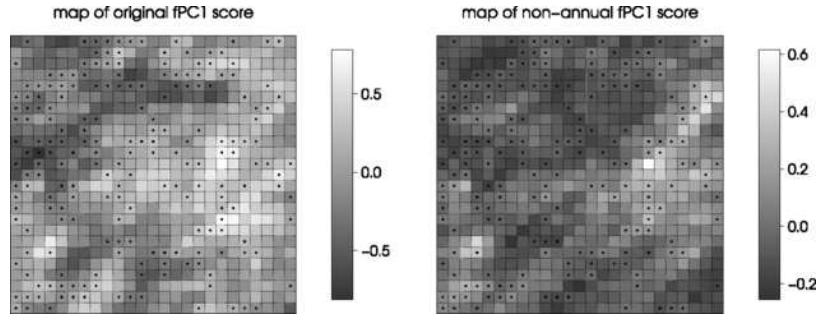}

\caption{Maps of component scores. The squares with dots in the middle
represent the pixels
we remove from the raw data set due to blocks of missing observations.
These two score maps are naturally oriented with north at the top.
Left: Projections of the first fPC onto EVI curves. Right: Projections
of the first nonannual fPC onto EVI curves. A clear south-west to
north-east pattern of correlation is evident in the nonannual fPC scores.}
\label{scoremap}
\end{figure}

Recall that our motivation of proposing PPCs is to separate annual and
nonannual variation in the EVI curves. We expect that change in variability,
if any, should be contained in the nonannual component. To uncover this
information, we look into the fPCs of this nonannual component. There are
distinct features in the first four nonannual fPCs; see Figure
\ref{non-annual}. In particular, a multiple of either the first
original or
nonannual fPC are added to and subtracted from the mean curve to facilitate
interpretation. The plus signs represent the curves which receive
positive fPC
scores, while the negative signs represent the curves which receive
negative fPC
scores. It is observed that the first nonannual fPC is mostly positive
in the
first three years and mostly negative in the last three years. The real message
of the first nonannual fPC is that the most dominant change of
variation is
the contrast of EVI relative to the cross-section mean between the
first three
years and the last three years. This contrast is also visualized by the gradual
change of relative positions of plus and minus signs, shown in the
upper panel
of the lower right plot in Figure~\ref{non-annual}. There are also
large peaks
during the 5th and 6th years, indicating events specific to
those years. The decreasing and last-two-year feature in the first nonannual
fPC strongly correspond to and enhance the features observed previously
in the
first and second original fPCs. Further, the second to the fourth nonannual
fPCs all capture information in particular years.

We can further investigate the spatial structure of the estimated aperiodic
effects by plotting the scores of the first nonannual fPC on a map of pixels.
The map on the right in Figure~\ref{scoremap} show a noticeable
south-west to north-east
correlation structure that may be indicative of local geographic features.
In preparing score maps in Figure~\ref{scoremap}, we imputed the
pixels which had been excluded
due to blocks of missing observations by using the functional
covariance structure estimated from the retained pixels. The imputation
procedure is discussed in detail in the supplemental article
[\citet{ppcsup}].
The existence of evident spatial\vadjust{\goodbreak}
correlation may require new approaches to fPCA. \citet
{pengandpaul} demonstrate
that fPCA remains consistent under mild assumptions on spatial correlation.
Alternatively, \citet{allen2011generalized} provide an approach
to directly
account for spatial correlation.\looseness=-1

%s3.7 #&#
\subsection{Tests of periodic variation}

The high correlation between the first few PPCs and associated
benchmarks gives rise to the question  of whether there is exact annual variation
contained in ${\boldsymbol\Gamma}_{46}$, the space of leading fPCs, or
PPCs (up to an orthogonal rotation). Note that the
first PPC has the highest correlation with any linear combination of
the annual
basis. So the test of whether there is exactly annual variation
contained in
${\boldsymbol\Gamma}_{46}$ is equivalent to testing the following hypothesis,
\begin{eqnarray*}
\mathbf{H}_0\dvtx  \rho_1 &=& 1,\\
\mathbf{H}_1\dvtx  \rho_1 &<& 1,
\end{eqnarray*}
where $\rho_1$ is the correlation between the first PPC and its
corresponding benchmark defined in \eqref{rho}. {Note we have two ways
to formulate this null hypothesis in terms of how we describe the
leading fPC subspace ${\boldsymbol\Gamma}_{46}$, either by the number
of fPCs spanning it, or the percentage of variation it explains. We
explore both formulations in the following analysis.}

 This null hypothesis does
not follow the classical test of correlation coefficients in a
multivariate setting
[see, e.g., \citet{mardia1980multivariate}]. Here we test that
the leading
principal components have a nontrivial intersection with a predefined subspace
rather than the independence of pairs of linear combinations of two random
vectors. To do so, we need to generate a null distribution for $\rho
_1$ which is no
longer invariant to the covariance under the null. We therefore seek an
approximate least-favorable covariance by a minimal perturbation of the
data so
as to satisfy $H_0$ and then apply a~bootstrap.

We first generate hypothesized curves to approximate the functional
covariance under the null hypothesis based
on curves ${\tilde z}_i(t)$'s.
We rewrite \eqref{decompose} as\looseness=-1
\begin{eqnarray*}
{\tilde z}_i(t) = s^{\xi}_{i1}\xi_1(t) + \sum_{j=2}^{M}s^{\xi
}_{ij}\xi
_j(t) +
\sum_{j=M+1}^{K}s^{\gamma}_{ij}\gamma_j(t), \qquad1\leq i\leq N.
\end{eqnarray*}\looseness=0
Under the null hypothesis, the first PPC and the first benchmark should
be identical. Then we can replace the first PPC with its associated
benchmark in the above equation and further write
%e7 #&#
%
\begin{eqnarray}
\label{replacenull}
{\bar z}_i(t) = s^{\xi}_{i1}\theta_1(t) + \sum_{j=2}^{M}s^{\xi
}_{ij}\xi
_j(t) +
\sum_{k=M+1}^{K}s^{\gamma}_{ij}\gamma_j(t), \qquad1\leq i\leq N.
\end{eqnarray}
${\bar z}_i(t)$'s are called hypothesized curves under replacement.
The eigenstructure of the covariance contained in ${\bar z}_i(t)$'s is
an approximated least-favorable eigenstructure\vadjust{\goodbreak} under the null. The
correlation between the first PPC and its benchmark of ${\bar
z}_i(t)$'s is 0.9999984
under both formulations of the null hypothesis, which we view as
sufficiently close to 1.
A distribution of the test statistic $\rho_1$ can now be generated
based on this
approximated null. One approach to obtaining a null distribution is to
assume a distribution on component
scores $s^{\xi}_{ij}$ and $s^{\gamma}_{ij}$, and produce a Monte Carlo
distribution of $\rho_1$. Here, we make no distributional assumptions
and apply a bootstrap procedure instead.

%f7 #&#
%
\begin{figure}

\includegraphics{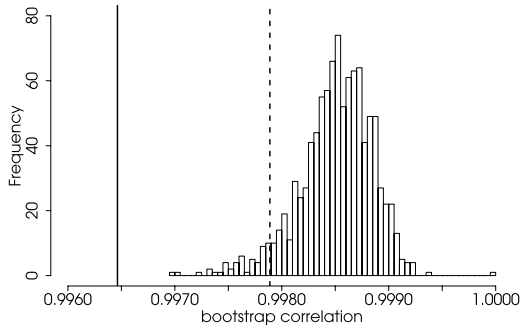}

\caption{Histogram of the first correlation $\rho_1$ derived from
bootstrap observations which are the sum of
null curves bootstrapped from ${\bar z}_i(t)$'s and bootstrap residuals.
The solid line is $\rho_1$ corresponding to the original curves of
Harvard Forest data. The dashed line is the
lower 0.05 critical value of the bootstrap distribution.}
\label{bootcor}
\end{figure}

We first sample with replacement from ${\bar z}_i(t)$'s to form
bootstrap null curves. In order to accommodate the effect of
pre-smoothing, we bootstrap residuals obtained from pre-smoothing
and add them onto each bootstrapped null curve. Then we re-smooth these
bootstrap observations and compute PPCs and
first correlations. This testing procedure follows the same framework
as that described in \citet{Li2011}, where the authors tested the
equality of functional means and covariances. Details of this procedure
are provided in the supplemental article [\citet{ppcsup}].
%Moreover, supplement file
%includes a simulation study where we examine the empirical size and
%power curve
%of this bootstrap procedure on a simulated data less complex than
%Harvard
%Forest data.

The histogram of bootstrap correlations with fixed number of fPCs is
shown in
Figure~\ref{bootcor}. The correlation between the first PPC and its associated
benchmark computed from the observed data is around 0.9965, which lies
at the
left tail of the bootstrap distribution, suggesting the major sources of
variation do not cover the strictly periodic functions.
%we should reject the null
%hypothesis of $\rho_1 = 1$.
 We can also apply this test to examine a fixed percentage of variation
explained instead of number of fPCs retained. The histogram of the null
distribution from this test is very similar and the null hypothesis is also
rejected. Readers are referred to the supplemental article [\citet
{ppcsup}] for detailed results.
% The
%histogram derived with fixed percentage of variation explained is
%slightly
%shifted to the left of the histogram derived with fixing the number of
%fPCs.
%However, it also leads to the rejection of the null hypothesis for the
%$\rho_1$
%calculated from the Harvard Forest data. This test shows that
%$\bf\boldsymbol\Gamma_{46}$ does not span anything in $\bf F_{46}$ and
%non-annual
%variation plays an important part in our data.

  We have viewed the null hypothesis derived by replacing the first PPC
with the first benchmark as being sufficiently close to the null
hypothesis for
our purposes. However, when this is not the case, the empirical first PPC
correlation\vadjust{\goodbreak} can be brought closer to 1 by inflating the first PPC
scores. This
method rescales the component score $s^{\xi}_{ij}$ in \eqref
{replacenull} and keeps $s^{\gamma}_{ij}$
fixed. This procedure allows\vspace*{1pt} the strength of annual signals to be
increased.
Rescaled curves are expected to have $\rho_1$ enlarged toward 1.
However, this
rescaling should be done in a way that distorts ${\bar z}_i(t)$'s and
the covariance kernel implied as little as possible. Hence, we put a
penalty on the
deviation of the hypothesized covariance kernel from the covariance kernel
computed from~${\bar z}_i(t)$'s, then we solve an optimization problem which
finds a balance between approximating the null hypothesis and
controlling for
divergence. Generally, let $\boldsymbol\tau= (\tau_1,\tau_2,\ldots
,\tau
_M)$ be the
rescaling vector. Then define hypothesized curves as\looseness=-1
%e8 #&#
%
\begin{eqnarray}
\label{inflatedcurve}
 \quad {\check z}_i(t, \boldsymbol\tau) = \tau_1s^{\xi}_{i1}\theta_1(t) +
\sum
_{j=2}^{M}\tau_js^{\xi}_{ij}\xi_j(t) +
\sum_{k=M+1}^{K}s^{\gamma}_{ij}\gamma_j(t), \qquad1\leq i\leq N.
\end{eqnarray}\looseness=0
The covariance kernels under replacement and inflation are given by
\begin{eqnarray*}
\boldsymbol\Omega(s,t) &=& \lambda^{\theta}_1\theta_1(s)\theta
_1(t) +
\sum_{j=2}^M\lambda^{\xi}_j\xi(s)\xi(t) +\sum_{j= M+1}^K\lambda
^{\gamma
}_j\gamma(s)\gamma(t),\\
\boldsymbol\Omega_0(s,t,\boldsymbol\tau) &=& \tau_1^2\lambda
^{\theta
}_1\theta_1(s)\theta_1(t) + \sum_{j=2}^M\tau_j^2\lambda^{\xi
}_j\xi(s)\xi
(t) + \sum_{j=M+1}^K\lambda^{\gamma}_j\gamma(s)\gamma(t),
\end{eqnarray*}
where $\boldsymbol\Omega(s,t)$ is the kernel based on curves under
replacement and
$\boldsymbol\Omega_0(s,t,\boldsymbol\tau)$ is the hypothesized kernel
based on rescaled curves ${\check z}_i(t, \boldsymbol\tau)$'s. {Under
the null hypothesis, $\theta_1(t),\{\xi_j(t)\}_{j = 2}^M$ and $\{
\gamma
_j(t)\}_{j = M + 1}^K$ are orthogonal to each other.} It can be shown
that the
Kullback--Leibler divergence of $\boldsymbol\Omega_0(s,t,\boldsymbol
\tau
)$ from~$\boldsymbol\Omega(s,t)$ is
given by
\begin{eqnarray*}
KL(\boldsymbol\Omega_0,\boldsymbol\Omega) = \frac{1}{2}\sum
_{j=1}^M(\tau_j^2 - 1 - \log\tau_j^2).
\end{eqnarray*}
Given ${\check z}_i(t,\boldsymbol\tau)$'s which are functions
of $\boldsymbol\tau$, we can compute PPCs and the first correlation
$\check{\rho}_1(\boldsymbol\tau)$.
Ideally, we want to minimize
$KL(\boldsymbol\Omega_0,\boldsymbol\Omega)$ with the restriction that
$\check{\rho}_1(\boldsymbol\tau) = 1$. This is achieved approximately
by placing a
large penalty on the difference between $\check{\rho}_1(\boldsymbol
\tau
)$ and 1. Then
we solve the following optimization:
%e9 #&#
%
\begin{eqnarray}
\label{inflatednull}
\min_{\boldsymbol\tau} KL(\boldsymbol\Omega_0,\boldsymbol\Omega) -
\lambda\log\check{\rho}_1(\boldsymbol\tau),
\end{eqnarray}
where $\lambda$ is a very large number.
Denote the optimizer to \eqref{inflatednull} by $\hat{\boldsymbol
\tau
}$. Then, ${\check z}_i(t,\hat{\boldsymbol\tau})$'s are constructed according to
\eqref{inflatedcurve}.
The eigenstructure implied by ${\check z}_i(t,\hat{\boldsymbol\tau})$'s
is closer
to the null hypothesis than that implied by ${\bar z}_i(t)$'s.

This procedure is investigated in detail in the supplemental article
[\citet{ppcsup}] where \eqref{inflatednull} is solved with a
sequence of $\lambda$ values. While the first correlation obtained by
this method increases, there is little effect on the test results.

%s4 #&#
\section{Sampling properties of PPC}\label{secsim}

In this section we explore the stability
and accuracy of PPC under random sampling. Two simulation schemes show the
sensitivity and robustness of PPC in identifying annual variation.

%s4.1 #&#
\subsection{Sensitivity}

In this simulation scheme, we demonstrate how sensitive the PPC is in detecting
annual variation. In the construction of the simulated curves, we take
the linear
combination of Fourier basis functions with different frequencies. We
create 6
sets of simulated curves. Each set contains 200 curves and incorporates a
different amount of annual variation by rescaling the coefficient of Fourier
basis functions which are annual. In particular, denote the $i$th curve
in the
$j$th set by $a^j_i(t)$. These curves are generated as a linear
combination of
longer term components and annual components as follows:
%e10 #&#
%
\begin{eqnarray}\label{sim1}
a^j_i(t) &=& \sum_{k=1}^3\sigma_{kji1}{\sin(k\omega t)} + \sum
_{k=1}^3\sigma_{kji2}{\cos(k\omega t)}
\nonumber
\\[-8pt]
\\[-8pt]&&{}+ \sqrt{L_j}\bigl(\sigma_{4ji1}{\sin(4\omega t)} + \sigma_{4ji2}{\cos
(4\omega t)}\bigr),
\nonumber
\end{eqnarray}
where $i = 1,2,\ldots,200$, $j = 1,2,\ldots,6$, $\omega= 2\pi/T$,
$\sigma_{kjil}\sim{\cal N}(0,1)$, i.i.d.,
$l = 1,2$, $L_1 = 0$, $L_2 = 0.6$, $L_3 = 0.8$, $L_4 = 1$, $L_5 = 1.1$,
and $L_6 = 1.3$.

\begin{figure}

\includegraphics{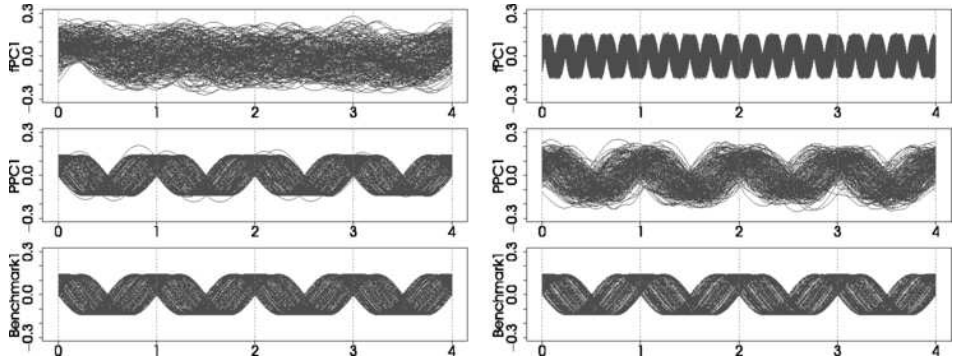}

\caption{Simulation results: estimated fPCs, PPCs and benchmarks.
Left: Simulation scheme 1 with
$L_4 = 1$; Right: Simulation scheme 2 with $L_3 = 5$.}
\label{simppc}
\end{figure}

$T$ is the time span of the simulated curves. We take $T = 100$ and $a^j_i(t)$
spans over 4 years. Thus, $\sin(4\omega t)$ and $\cos(4\omega t)$ are sources
of annual variation. The Fourier basis functions in the first two
components of
$\eqref{sim1}$ are orthogonal to annual basis functions and thus do not
contribute to the annual variation. The $L_j$'s control the amount of annual
variation. The larger the $L_j$, the greater the amount of annual
variation. We
compute PPCs with $80\%$ of total variation cutoff in choosing how many fPCs
we retain in all 6 sets. The result for $L_4 = 1$ is shown in the left
3 plots
of Figure~\ref{simppc}. The fPCs do not capture the underlying source of
annual variation.
%f8 #&#
%

How much each sinusoidal function is reflected in retained fPCs depends
on both
the sample variance and covariance of $\sigma_{k4il}$ and on their interaction
with other sources of variation. However, $\sin(4\omega t)$ and $\cos
(4\omega
t)$ can be identified by PPCs even when their variation are on the same level
($L_4 = 1$) as other sources. The benchmarks exactly reproduce the annual
signals, however, with phase shifting. The shifted phase is caused by the
randomness in sampling $\sigma_{44i1}$ and $\sigma_{44i2}$.

\begin{figure}

\includegraphics{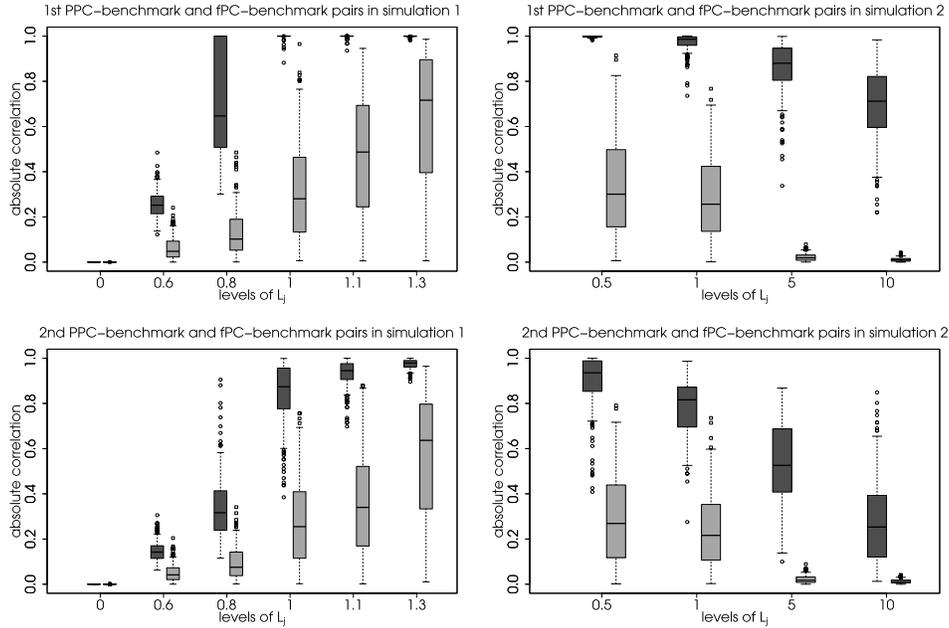}

\caption{Simulation results: periodicity of estimated PPCs. Dark boxes
are correlations between
PPCs, $\xi_j$ and associated benchmarks $\theta_j$.
Light boxes are correlations between fPCs, $\gamma_j$ and associated
benchmarks $\theta_j$.
Upper left: Simulation scheme 1 results on $\xi_1$, $\gamma_1$ and~$\theta_1$;
Upper right: Simulation scheme 2 results on $\xi_1$, $\gamma_1$ and~$\theta_1$;
Lower left: Simulation scheme~1 results on $\xi_2$, $\gamma_2$ and~$\theta_2$;
Lower right: Simulation scheme 2 results on $\xi_2$, $\gamma_2$ and~$\theta_2$.}
\label{simcor}
\end{figure}

To summarize the simulation results for all $L_j$'s, we compute the
standard\-ized-inner-product (correlation) between the PPC-benchmark pair and
between the fPC-benchmark pair. Since the sign is irrelevant with both
fPCs and
PPCs, we take the absolute values of the correlations. The boxplot of the
unsigned correlations of the first and the second pairs are shown in the
upper-left and lower-left plots in Figure~\ref{simcor}. For both the
first and
second pairs,\vadjust{\goodbreak} fPC-benchmark correlations show an increasing trend
toward 1. As
we include more annual variation, the fPCs will tend to be more nearly annual.
However, the speed of fPC-benchmark correlations going to 1 is much slower
compared to that of PPC-benchmark correlations. Moreover, PPC-benchmark
correlations are always higher than fPC-benchmark correlations for all $L_j$'s.
This observation demonstrates the sensitivity of PPCs in detecting annual
variation among curves.

%f9 #&#
%

%s4.2 #&#
\subsection{Robustness}

In the second simulation scheme, we add one more source of variation
which is
generated by nonannual Fourier basis functions with high frequency. We
call it
high frequency disturbance (HFD). According to the definition, the HFD
is not a
source of annual variation. In our simulation study, we construct 4
sets of
simulated data, 200 curves each, which contain different levels of HFD.
We test
PPCs' robustness of detecting annual variation in the presence of HFD.
Specifically, denote the $i$th curve in the $j$th set by $b^j_i(t)$.
Then it is
generated as
\begin{eqnarray*}
b^j_i(t) &=& \sum_{k=1}^4\sigma_{kji1}{\sin(k\omega t)} + \sum
_{k=1}^4\sigma_{kji2}{\cos(k\omega t)}\\
&&{} +
\sqrt{L_j}\bigl(\sigma_{zji1}{\sin(z\omega t)} + \sigma_{zji2}{\cos
(z\omega t)}\bigr),
\end{eqnarray*}
where $z = 19$, $i = 1,2,\ldots,200$, $j = 1,2,3,4$, $\omega=
2\pi/T$,
$\sigma_{\cdot jil}\sim{\cal N}(0,1)$, i.i.d., $l = 1,2$, $L_1 =
0.5$, $L_2
= 1$, $L_3 = 5$, and $L_4 = 10$. $T$ equals 100, spanning over 4 years, as
in the first simulation. The functions $\sin(4\omega t)$ and $\cos
(4\omega t)$ are still
the sources of annual variation which have the same amount of variation
in the 4~sets of this simulation scheme. $z$ is the frequency of HFD and is set
to be 19
in our simulation. $\sin(z\omega t)$ and $\cos(z\omega t)$ are HFD whose
amount of variation varies and are controlled by $L_j$'s. Larger $L_j$ value
suggests greater amount of HFD and, hence, it is more difficult to
extract annual
signals for larger $L_j$'s.\vadjust{\goodbreak} In this scheme, we also use $80\%$ as the cutoff
to decide the number of fPCs we retain. The computed PPCs for $L_3 = 5$ is
shown in Figure~\ref{simppc}. With amount of HFD
5 times
as great as annual variation, the fPCs are dominated by HFD and thus
show a
clear 19-periodic pattern. However, our first two PPCs still show a reasonably
good annual pattern. To summarize results for all $L_j$'s, we plot the
fPC-benchmark and PPC-benchmark correlations of the first two pairs in the
upper-right and lower-right plots in Figure~\ref{simcor}. Again, for both
pairs, the fPC-benchmark correlations are always lower than the PPC-benchmark
correlations. Further, even for large HFD contamination ($L_j \ge5$)
when the
fPC-benchmark correlations hover near zero, the PPC-benchmark correlations
display much higher values, suggesting that the PPCs provide more robust
directions compared to fPCs as the amount of HFD increases.

Based on these two simulations, we find PPCs are both sensitive and robust
identifiers of the source of annual variation.

%s5 #&#
\section{Conclusion}

Despite the popularity of functional principal component analysis, little
attention has been paid to the problem of factor rotation to improve the\vadjust{\goodbreak}
interpretability of modeled principal component directions. The
smoothness, or
ordering, properties of functional data analysis mean that factor rotation
methods that are applicable for multivariate data are not always
appropriate in
a functional context. Conversely, new factor rotation methods may be applicable
in functional data analysis that do not have analogues in multivariate
statistics. As for all
factor rotation methods, it is important to recall that the resulting
directions are obtained as an interpretable means of representing the data,
rather than independent mechanistic sources of variance.

In this paper, we have presented a factor rotation method motivated by remote
sensing data and intended to improve our understanding of factors
involved in
ecological responses to climate change. In this data set we seek to
differentiate seasonal sources of variation from both longer-term and localized
effects. To do this, we present principal periodic components as a~means of
extracting nearly-periodic directions in the data. This factor rotation
has the
advantage of being efficiently implementable via canonical correlation
analysis and effective at extracting periodic information. We have developed
graphical tools to assess the level of periodicity in the data and to
decide on
thresholds between periodic and aperiodic signals. Further, a~heuristic test
of exact periodicity demonstrates that the addition of some further
flexibility in our periodic signals is appropriate.

At its most general, our approach can be described as a rotation
toward an interpretable subspace and applies to multivariate factor rotation
as well as in functional data analysis. In our application, the set of periodic
functions represents the most clearly relevant subspace for interpretation.
However, alternative subspaces may be useful in other contexts; for
example, in
\citet{koulis} a psychological experiment is described in which a
stimulus is
changed at prespecified times and a data-set of continuously-measured
responses is recorded.
In this case, a basis of step functions corresponding to
change-times represents a relevant reference subspace with which to
examine the
functional response to the stimulus sequence. The choice of reference subspace
depends strongly on the details of the application at hand.
In our own application, we could have sought further rotations of aperiodic
signals toward linear or exponential trends as a means of separating long-term
effects from effects localized to individual years. Beyond this
approach, we
expect a more general exploration of sources of variation within the
context of functional data analysis to be an important source of future
research directions.

\begin{supplement}%[id=suppA]
\stitle{Description of data and details of simulation\\}
\slink[doi,text={10.1214/11-\break
AOAS518SUPP}]{10.1214/11-AOAS518SUPP} %[doi,text={...}] - jei reikia
%suskaldyti doi
\slink[url]{http://lib.stat.cmu.edu/aoas/518/supplement.pdf}
\sdatatype{.pdf}
\sdescription{The supplementary material is divided into 3 sections.
The first section provides a detailed description of the Harvard Forest\vadjust{\goodbreak}
data that is used in this article, including preprocessing steps. We
also provide a detailed description of the imputation steps for pixels
with missing observations. The second section provides a description of
\textit{Annual Information} and its application is demonstrated through
a simulation study. The last section provides results related to the
bootstrap hypothesis testing procedure proposed in this article. In
particular, we present the test results on the Harvard Forest data and
simulation studies where we explore the empirical power curve and size
on simulated data sets.}
\end{supplement}

% imsref loaded by smiklovaite, 2012-01-09 20:19:39
% imsref loaded by smiklovaite, 2012-01-10 08:31:58
%

\printaddresses

\end{document}